\documentclass[12pt, amsfonts, amssymb]{article}

\usepackage{amssymb}
\usepackage{amsmath}
\usepackage{latexsym}

\def\be{\begin{equation}}
\def\ee{\end{equation}}
\def\lb{\left(}
\def\rb{\right)}

\begin{document}

\begin{titlepage}
 \vskip 1cm

\begin{flushright}
CALT-68-2459
\end{flushright}

\baselineskip=20pt

\centerline{\Large {\bf {Monopole Operators in
Three-Dimensional}}} \centerline{\Large \bf $\bf{\cal N}=4$ SYM
and Mirror Symmetry}

\vskip 0.5cm

\renewcommand{\thefootnote}{\fnsymbol{footnote}}

\centerline{Vadim Borokhov$^*$\footnotetext{\hspace*{-17pt}${}^*$
 borokhov@theory.caltech.edu}}

\vskip 0.5cm

\centerline{{ \it California Institute of Technology }}
\centerline{{ \it Pasadena, CA 91125, USA }}

\vskip 1cm \centerline{\sc {\bf Abstract}} \vskip 0.5 cm

We study non-abelian monopole operators in the infrared limit of
three-dimensional $SU(N_c)$ and ${\cal N}=4$ $SU(2)$ gauge
theories. Using large $N_f$ expansion and operator-state
isomorphism of the resulting superconformal field theories, we
construct monopole operators which are (anti-)chiral primaries and
compute their charges under the global symmetries. Predictions of
three-dimensional mirror symmetry for the quantum numbers of these
monopole operators are verified.

\vskip .5in

\end{titlepage}

\section{\Large {\bf Introduction}}

A non-perturbative duality in three dimensions is known as 3d
mirror symmetry. It was first proposed by K. Intriligator and N.
Seiberg in Ref.\cite{IS} and further studied in
Refs.\cite{deBHOO}-\cite{GT}. The mirror symmetry predicts quantum
equivalence of two different theories in the IR limit. In this
regime a supersymmetric gauge theory is described by a strongly
coupled superconformal field theory. The duality exchanges masses
and Fayet-Iliopoulos terms as well as the Coulomb and Higgs
branches implying that electrically charged particles in one
theory correspond to the magnetically charged objects (monopoles)
in the other. Also, since the Higgs branch does not receive
quantum corrections and the Coulomb branch does, mirror symmetry
exchanges classical  effects in one theory with quantum effects in
the dual theory. Many aspects of three-dimensional mirror symmetry
have a string theory origin.

This paper extends the analysis of Refs.\cite{BKW1}-\cite{BKW2} to
the non-abelian gauge theories. We study monopole operators in the
IR limit of $SU(N_c)$ non-supersymmetric  Yang-Mills theories as
well as ${\cal N}=4$
 $SU(2)$ supersymmetric Yang-Mills models with large number of flavors
$N_f$. Conformal weight of a generic monopole operator in
non-supersymmetric gauge theory is irrational. On the other hand,
supersymmetric gauge theories have monopole operators which are
superconformal (anti-)chiral primaries. The conformal dimensions
of such operators are uniquely determined by their $R$-symmetry
representations. The $R$-symmetry group of ${\cal N}=4$
supersymmetric theory is given by $SU(2)\times SU(2)$ and
conformal dimensions of the (anti-)chiral primary operators are
integral. The mirror symmetry predicts the spectrum and quantum
numbers of (anti-)chiral primary operators including the ones with
magnetic charges.
 We use $1/N_f$ expansion and the operator-state isomorphism of the
resulting conformal field theories to study transformation
properties of  monopole operators under the global symmetries and
verify the mirror symmetry predictions.

In Refs.\cite{JT}-\cite{Templeton} it was demonstrated that
three-dimensional gauge theories have severe perturbative infrared
divergences due to logarithms of the coupling constant. In
Ref.\cite{AN} it was shown that for three-dimensional QCD, the
$1/N_f$ expansion can be defined in such a way that the infrared
divergences are absent in each order of
 the expansion and the theory
has  IR fixed point. For large $N_f$ the non-abelian interactions
of gluons are suppressed and dynamics of the theory becomes
similar to that of an abelian theory. In Refs.\cite{AN}-\cite{DSS}
it is claimed that for $N_f$ smaller than a certain critical value
the dynamical fermion mass is generated. These conclusions were
supported by the lattice simulations in Ref.\cite{DKK}. The phase
transition takes place at finite $N_f$ and does not affect the
dynamics at large $N_f$ which is studied in this paper. However,
it indicates that $1/N_f$ expansion has a finite radius of
convergence. In the case of ${\cal N}=4$ supersymmetric Yang-Mills
theories, IR limit of the theory is given by an interacting
superconformal field theory and there is no evidence of the phase
transitions at finite $N_f$. It is possible that $1/N_f$ expansion
is convergent all  the way down to $N_f=1$. Our analysis is
performed in the origin of the moduli space, thus extending
results of Ref.\cite{five}, where implications of the mirror
symmetry have been verified on the Coulomb branch of ${\cal N}=2$
supersymmetric Yang-Mills theories.

The paper is organized as follows. In Section~\ref{MonopolesinYM}
we study monopole operators in the IR limit of $SU(N_c)$ gauge
theories and determine their quantum numbers at large $N_f$.
Monopole operators in the IR limit of ${\cal N}=4$ $SU(2)$ gauge
 theories are considered in Section~\ref{MonopolesinSYM}. We
discuss our results in Section~\ref{Discussion}.

\bigskip

\section{\Large \bf Monopole Operators in $\bf SU(N_c)$ Gauge
Theories}\label{MonopolesinYM}

\subsection{\large \bf IR Limit of $\bf SU(N_c)$ Gauge
Theories}\label{SYM}

Consider a three-dimensional Eucledean Yang-Mills action for $N_f$
flavors of matter fermions in the fundamental representation of
the gauge group $SU(N_c)$ with generators
$\{T^\alpha\}$\footnote{We use $\textrm{Tr}(T^\alpha
T^\beta)=\frac12\delta^{\alpha\beta}$ normalization.},
($\alpha=1,\dots,N_c^2-1$): \be\label{YM}
S=\int\phantom{}d^3x\lb\frac{1}{4e^{2}}\textrm{Tr
}{V_{ij}V^{ij}}+i\sum_{s=1}^{N_f}
\bar{\psi}^s\vec{\sigma}\lb\vec{\nabla}+i\vec{V}\rb\psi^s\rb, \ee
where $\psi$ are complex two-component spinors,
$\vec{V}=\vec{V}^{\alpha} T^\alpha$ is gauge potential with a
field-strength $V_{ij}=\lb
\partial_i V^\alpha_j-\partial_j
V^\alpha_i-f^{\alpha\beta\gamma}V^\beta_i V^\gamma_j \rb
T^\alpha$, $f^{\alpha\beta\gamma}$ are the structure constants. To
avoid a parity anomaly, Ref.\cite{R}, we choose $N_f$ to be even.
Action (\ref{YM}) is invariant under the flavor symmetry
$U(N_f)_{flavor}$.

There are two ways to classify monopoles in non-abelian theories.
A dynamical description of monopoles in terms of weight vectors of
the dual of (unbroken) gauge group was developed by Goddard,
Nuyts, and Olive (GNO) in Ref.\cite{GNO}; topological
classification in terms of $\pi_1$ was suggested by Lubkin in
Ref.\cite{L}, (see also Ref.\cite{C} for a review). It is well
known that in $R^{1,3}$ the dynamical (GNO) monopoles with
vanishing topological charges are unstable in the small coupling
limit. We will study the dynamical monopoles in the IR limit of
(\ref{YM}). The theory is free in the UV limit
($\frac{e^2}{\Lambda}\to0$, where $\Lambda$ is a renormalization
scale) and is strongly coupled in the IR
($\frac{e^2}{\Lambda}\to\infty$). In the strong coupling regime
the dominant contribution to the gauge field effective action is
given by the matter fields and stability analysis of GNO monopoles
performed at weak coupling is no longer applicable. Since matter
fields belong to the fundamental representation, the effective
gauge group is given by $SU(N_c)$. The corresponding $\pi_1$ is
trivial and all dynamical monopoles have vanishing topological
charges. The GNO monopoles of $SU(N_c)$ are given by
\be\label{GNOMonopole} V^N=H(1-\cos{\theta})d\varphi,\quad
V^S=-H(1+\cos{\theta})d\varphi, \ee where $V^N$ and $V^S$
correspond to gauge potentials on upper and lower hemispheres
respectively. $H$ is a constant traceless hermitian $N_c\times
N_c$ matrix, which can be assumed to be diagonal. On the equator
$V^N$ and $V^S$ are transformed into each other by a gauge
transformation with a group element $\exp{\lb2iH\varphi\rb}$. This
transformation is single-valued if \be\label{H}
H=\frac12\textrm{diag}(q_1,q_2,\dots,q_{N_c}) \ee with integers
$q_a$, ($a=1,..,N_c$), \be\label{vanishingtrace}
\sum_{a=1}^{N_c}q_a=0. \ee Consider a path integral over matter
and gauge fields on the punctured $R^3$. Integration over the
gauge fields asymptotically approaching (\ref{GNOMonopole}) at the
removed point of $R^3$ and corresponds to an insertion of a
topology-changing operator with magnetic charge $H$. To complete
definition of the topology-changing operator we have to specify
the behavior of matter fields at the insertion point. Thus
topology-changing operators with a given magnetic charge are
classified by the behavior of the matter fields near the
singularity. In the IR limit the theory (\ref{YM}) flows to the
interacting conformal field theory (CFT). In three-dimensional CFT
operators on $R^3$ are in one-to-one correspondence with
normalizable states on $S^2\times R$. Namely, insertion of a
topology-changing operator in the origin of $R^3$ corresponds to a
certain in-going state in the radially quantized theory on
$S^2\times R$. Hamiltonian of the radially quantized theory is
identical to the dilatation operator on $R^3$.  In unitary CFT all
physical operators including topology-changing ones are classified
by the lowest-weight irreducible representations labelled by the
primary operators. We will say that topology-changing operator is
a monopole operator, i.e., corresponds to the creation of the GNO
monopole, if such an operator has the lowest conformal weight
among the topology-changing operators with a given magnetic charge
$H$. Since conformal transformations do not affect the magnetic
charge, the monopole operators are conformal primaries. Our task
is to determine spin, conformal weight and other quantum numbers
of the monopole operators.

In the IR limit kinetic term for the gauge field can be neglected
and integration over matter fields produces effective action for
the gauge field proportional to $N_f$. Although IR theory is
strongly coupled, the effective Planck constant is given by
$1/N_f$ and in the large $N_f$ limit the CFT becomes weakly
coupled.  It is natural to assume that saddle point of the gauge
field effective action is invariant under rotations and
corresponds to the GNO monopole.  Since fluctuations of the gauge
field are suppressed, it can be treated as a classical background.
Thus, in the large $N_f$ limit we have matter fermions moving in a
presence of the GNO monopole. Therefore, a monopole operator is
mapped to a Fock vacuum for matter fields moving in a monopole
background on $S^2\times R$. Conformal weight of the monopole
operator is equal to Casimir energy of the corresponding vacuum
state relative to the vacuum state with vanishing monopole charge.

\subsection{\large \bf Radial Quantization}

Let us implement the procedure outlined in the previous section.
Namely, we consider CFT which appears in the IR limit of the
theory (\ref{YM}). We neglect the kinetic term of a gauge field,
introduce a radial time variable $\tau=\ln{r}$ and perform the
Weyl rescaling to obtain metric on $S^2\times R$:
$$
ds^2=d\tau^2+d\theta^2+\sin^2{\theta} d\varphi^2.
$$
 Since a gauge potential of the GNO monopole (\ref{GNOMonopole}) with
$H$ given by Eqs.(\ref{H})-(\ref{vanishingtrace}) is diagonal in
color indices we may use results of Ref.\cite{BKW1}\footnote{$q$
in present paper equals twice that in Ref.\cite{BKW1}.} for
fermionic energy spectra on $S^2\times R$. We conclude that for
each $\psi^s_a$, where $s=1,\dots,N_f$ and $a=1,\dots,N_c$ are
flavor and color indices respectively, the energy spectrum is
given by
$$
E_n=\pm\sqrt{|q_a|n+n^2},\quad n=1,2,\dots
$$
Each energy mode has a degeneracy $2|E_n|$ and spin
$j=|E_n|-\frac12$. In addition, there are $|q_a|$ zero-energy
modes which transform as an irreducible representation of the
rotation group $SU(2)_{rot}$ with spin $j=\frac12\lb|q_a|-1\rb$.
In the large $N_f$ limit leading contribution to the conformal
weight $h_{\{q\}}$ of the GNO $SU(N_c)$ monopole is given by
$$
h_{\{q\}}=N_f\sum_{a=1}^{N_c}\lb\frac16\sqrt{1+|q_a|}\lb|q_a|-2\rb+\phantom{1111111111111111111111111111111111111}\right.
$$
$$
+\left.4\textrm{Im}\int_0^\infty\phantom{} dt\left[\lb
it+\frac{|q_a|}{2}+1\rb \sqrt{\lb
it+\frac{|q_a|}{2}+1\rb^2-\frac{q_a^2}{4}} \right]\frac{1}{e^{2\pi
t}-1} \rb,
$$
where branch of the square root under the integral is the one
which is positive on the positive real axis.

Let us specialize in the case of GNO monopole with minimum
magnetic charge: \be\label{FundamentalMonopole}
H=\frac12(1,-1,0,\dots,0), \ee and denote the fermionic
nonzero-energy mode annihilation operators by $a^{s}_{akm}$,
$b^{s}_{akm}$, where $k$ labels the energy level, and $m$ accounts
for a degeneracy. Fermionic zero-energy modes have vanishing spin
and are present for $\psi^s_1$ and $\psi^s_2$ only. The
corresponding annihilation operators we denote as $c^{s}_1$ and
$c^s_2$. Consider a Fock space of states obtained by acting with
creation operators on a state $|vac\rangle$, which is defined as a
state annihilated by all the annihilation operators. Those
elements of the Fock space which satisfy the Gauss-law constraints
form the physical Fock space.

The background (\ref{GNOMonopole}) with $H$ given by
Eq.(\ref{FundamentalMonopole}) breaks gauge group $G=SU(N_c)$ to
$\bar{G}=U(1)$ for $N_c=2$ and $\bar{G}=SU(N_c-2)\times U(1)\times
U(1)$ for $N_c>2$, where generators of the two $U(1)$ groups are
given by $\lb1,-1,0,\dots,0\rb$ and $\lb
2-N_c,2-N_c,2,\dots,2\rb$. Let $\bar{T}^{\bar{\alpha}}$ be
generators of $\bar{G}$. In a quantum theory we impose Gauss-law
constraints on physical states. In the IR limit it implies that
they are annihilated by the charge density operators
$\rho^{\bar{\alpha}}$. Consider charges $Q^{\bar{\alpha}}$
obtained by integration of $\rho^{\bar{\alpha}}$ over $S^2$. The
most general form of the corresponding quantum operators is
$$
Q^{\bar{\alpha}}=Q^{\bar{\alpha}}_++Q^{\bar{\alpha}}_0,
$$
where $Q^{\bar{\alpha}}_0$ denote all terms that act within a
zero-mode Fock space and $Q^{\bar{\alpha}}_+$ are assumed to be
normal-ordered. Using explicit form of zero-energy solutions we
find
$$
Q^{\bar{\alpha}}_0=c^{+s}_1\bar{T}^{\bar{\alpha}}_{11}c^s_1+c^{+s}_2\bar{T}^{\bar{\alpha}}_{22}c^s_2+n^{\bar{\alpha}},
$$
where C-numbers $n^{\bar{\alpha}}$ account for operator-ordering
ambiguities. Since  zero modes are rotationally invariant, the
Gauss-law constraints in the zero-mode Fock space are translated
into requirements that the states are annihilated by
$Q^{\bar{\alpha}}_0$.

In the case of $N_c=2$ we have
$$
Q_0=\frac12\lb c^{+s}_1c^s_1-c^{+s}_2c^s_2\rb+n.
$$
The zero-mode space in spanned by the $2^{2N_f}$ states
$$
|{vac} \rangle,\quad c^{+s_1}_1|{vac}\rangle, \quad
c^{+s_1}_2|{vac}\rangle,\quad \dots, \qquad c^{+s_1}_1\dots
c^{+s_{N_f}}_1c^{+s_1}_2\dots c^{+s_{N_f}}_2|{vac}\rangle.
$$
A Fock vacuum state as well as completely filled state have
$Q_0$-charge given by $n$. Since the monopole background is
invariant under $CP$ symmetry,  we require $CP$-invariance of the
$Q_0$ spectrum. Therefore, $n=0$ and we have the following
physical vacuum states transforming as scalars under $SU(2)_{rot}$
$$
|{vac}\rangle,\quad c^{+s_1}_1\dots c^{+s_l}_1c^{+p_1}_2\dots
c^{+p_l}_2|{vac}\rangle, \qquad l=1,\dots,N_f.
$$
Each set of the physical vacuum states labelled by $l$ transforms
as a product of two rank-$l$ antisymmetric tensor representations
under $U(N_f)_{flavor}$.

For $N_c>2$ we choose $\bar{T}^1$ and $\bar{T}^2$ to be generators
of the two $U(1)$ groups so that the only zero-mode contributions
are
$$
Q^1_0=\frac12\lb c^{+s}_1c^s_1-c^{+s}_2c^s_2\rb+n^1,\quad
Q^2_0=-\frac12\sqrt{\frac{N_c-2}{N_c}}\lb
c^{+s}_1c^s_1+c^{+s}_2c^s_2\rb+n^2.
$$
 In this case $CP$-invariance gives $n^1=0$ and
$n^2=\frac12\sqrt{\frac{N_c-2}{N_c}}N_f$. Therefore, we have ${N_f
\choose \frac12N_f}^2$ physical vacuum states
$$
c^{+s_1}_1\dots c^{+s_{N_f/2}}_1c^{+p_1}_2\dots
c^{+p_{N_f/2}}_2|{vac}\rangle,
$$
transforming as scalars under $SU(2)_{rot}$ and as a product of
two rank-$N_f/2$ antisymmetric tensor representations of
$U(N_f)_{flavor}$.

\bigskip

\section{\Large \bf Monopole Operators in $\bf {\cal N}=4$ $\bf SU(2)$
Gauge Theory}\label{MonopolesinSYM}

\subsection{\large \bf IR Limit of $\bf{{\cal N}=4}$ $\bf{SU(2)}$
Gauge Theory}

Consider three-dimensional Eucledean ${\cal N}=4$ supersymmetric
theory of vector multiplet ${\cal V}$ in the adjoint
representation of the gauge group $SU(2)_{gauge}$ and $N_f$ matter
hypermultiplets ${\cal Q}^s$, $(s=1,\dots,N_f)$, transforming
under the fundamental representation. The action in terms of
three-dimensional ${\cal N}=2$ superspace formalism is given in
the Appendix. Decompositions of ${\cal N}=4$ multiplets into
${\cal N}=2$ multiplets are given in the following table
\begin{displaymath}
\begin{tabular}{c|c}
\hline
${\cal N}=4$&${\cal N}=2$\\
\hline
&\\
\textrm{Vector multiplet }${\cal V}$& \textrm{Vector multiplet
}$V=(V_i,\chi, \lambda,\bar{\lambda}, D)$,\\
&\textrm{Chiral multiplet }$\Phi=(\phi,\eta,K)$.\\
&\\
\hline
&\\
\textrm{Hypermultiplet }${\cal Q}$&\textrm{Chiral multiplets
}$Q=(A,\psi,F)$,\\
&$\tilde{Q}=(\tilde{A}, \tilde{\psi},\tilde{F})$,\\
&\\
\hline
 \end{tabular}
\end{displaymath}
where $V_i$ is a vector field in the adjoint representation of the
gauge group, $\chi$ and $\phi$ are real and complex adjoint
scalars respectively, $\lambda$, $\bar{\lambda}$, and $\eta$ are
the gluinos, whereas fields $D$ and $K$ are auxiliary. Scalar $A$
($\tilde{A}$), spinor $\psi$ ($\tilde{\psi}$), and auxiliary field
$F$ ($\tilde{F}$) transform according to (anti-)fundamental
representation of the gauge group:
$$
Q\to e^{i\omega^\alpha T^\alpha} Q,\quad \tilde{Q}\to
\tilde{Q}e^{-i\omega^\alpha T^\alpha},
$$
under the gauge transformation with parameters $\omega^\alpha(x)$.
Since all representations of $SU(2)$ are pseudo-real, we may
define a chiral superfield
$$
\Psi^{a}=\frac{1}{\sqrt{2}}{ Q^{a}-\epsilon^{ab}\tilde{Q}_b
\choose i\left[ Q^a+\epsilon^{ab}\tilde{Q}_b\right]},
$$
where $\epsilon^{ab}$ is antisymmetric tensor with
$\epsilon^{12}=1$. Therefore kinetic term for a hypermultiplet has
the form
$$
\int{\phantom{1}} d^2\theta
d^2\bar{\theta}\sum_{I=1}^{2N_f}\Psi^+_I e^{2V} \Psi_I,
$$
where we used the identities
$$\epsilon_{ab}T^{\alpha b}_c\epsilon^{cd}=-(T^{\alpha a}_d)^T,\quad
\epsilon_{ab} \epsilon^{bc}=\delta^c_a.
$$
The superpotential is
$$
W=i\sqrt{2}\sum_{s=1}^{N_f}\tilde{Q}^s\Phi
Q^s=\frac{i}{\sqrt{2}}\sum_{I=1}^{2N_f}\Psi^a_I\epsilon_{ab}\Phi^b_c\Psi^c_I.
$$
The kinetic term is invariant under $SU(2N_f)$ flavor symmetry.
The superpotential, however, is invariant under $SO(2N_f)$
subgroup only.

$4N_f-6$ dimensional\footnote{Moduli space dimensions are assumed
to be complex.} Higgs branch is labelled by the mesons
$M_{IJ}=\Psi^a_I\epsilon_{ab}\Psi^b_J$. Using an identity
$$
\epsilon^{I_1\dots
I_{2N_f}}\Psi^{a}_{I_1}\Psi^{b}_{I_2}\Psi^{c}_{I_3}
\Psi^{d}_{I_4}=0,
$$
we obtain the constraints $\epsilon^{I_1\dots
I_{2N_f}}M_{I_1I_2}M_{I_3I_4}=0$.
 The F-flatness condition implies $M^2_{IJ}=0$.

On the Coulomb branch adjoint scalars $\chi$ and $\Phi$ can have
nonvanishing expectation values.  Let us make a gauge
transformation to obtain $\chi=\chi^{(3)}T^3$. Dualizing a photon
$V^{(3)}=*d\sigma^{(3)}$  we construct a chiral superfield
$\Upsilon=\chi^{(3)}+i\sigma^{(3)}+\dots$ Potential energy density
for scalars $\chi$ and $\Phi$ is given by
 $U=U_1+U_2$, with
$$
U_1\sim \textrm{Tr}\lb\left[\Phi,\Phi^+\right]\rb^2,\quad U_2\sim
\textrm{Tr}\lb\chi^2\rb \textrm{Tr}\lb
\Phi^+\Phi\rb-\left|\textrm{Tr}\lb\chi\Phi\rb\right|^2.
$$
Vanishing of the potential gives $\Phi=\Phi^{(3)}T^3$. Residual
gauge symmetries are $U(1)_{gauge}$ generated by $T^3$ and Weyl
subgroup $Z_2$ acting by
$(\Upsilon,\Phi^{(3)})\to(-\Upsilon,-\Phi^{(3)})$. Moreover, we
have $\Upsilon\sim\Upsilon+4\pi e^2i$. Let us introduce a pair of
operators $Y_+$ and $Y_-$ corresponding to positive and negative
expectation values of $\chi^{(3)}$ respectively. For large
positive (negative) $\chi^{(3)}$ we have $Y_+\sim
e^{\Upsilon/(2e^2)}$ ($Y_-\sim e^{-\Upsilon/(2e^2)}$). We
emphasize that none of the $Y_\pm$ is gauge invariant. In fact,
$Y_+\leftrightarrow Y_-$ under the Weyl subgroup $Z_2$.
 The gauge invariant coordinates on the Coulomb branch are
\be\label{CB} u=i{\lb Y_+-Y_-\rb}\Phi^{(3)},\quad v={\lb
Y_++Y_-\rb}, \qquad w=\lb\Phi^{(3)}\rb^2. \ee In a semiclassical
limit we have an equation \be\label{classCoulomb} u^2+v^2w=0. \ee
 Since the Coulomb branch receives quantum corrections we expect
modification of the Eq.(\ref{classCoulomb}).

Three-dimensional ${\cal N}=4$ theory has an $R$-symmetry group
given by $SU(2)_R\times SU(2)_N$. There are $SU(2)_R$ and
$SU(2)_N$ gluino doublets, scalars $A$ $(A^+)$ and $\tilde{A}^+$
$(\tilde{A})$ make a doublet of $SU(2)_R$ and are singlets of
$SU(2)_N$, spinors $\psi$ $(\bar{\psi})$ and $\bar{\tilde{\psi}}$
$(\tilde{\psi})$ transform as a doublet of $SU(2)_N$ and singlets
of $SU(2)_R$. Scalars $\chi$, $\phi$, and $\phi^+$ form a triplet
of $SU(2)_N$ and are neutral under $SU(2)_R$. In three-dimensional
${\cal N}=2$ superspace formalism only the maximal torus
$U(1)\times U(1)$ of the $R$-symmetry is manifest. Let us
introduce a set of manifest $R$-symmetries denoted as $U(1)_N$,
$U(1)_B$,   and $U(1)_R$ with the corresponding charges given in
the table
\begin{displaymath}
\begin{tabular}{c|c|c|c}
& $N$ & $B$ & $R$\\
\hline
$Q$ & $0$ & $1$ & $1/2$\\
\hline
$\tilde{Q}$ & $0$ & $1$ & $1/2$ \\
\hline
$\Phi$ & $2$ & $-2$ & $1$\\
\end{tabular}
\end{displaymath}
It is easy to see that $B$-charge of the Grassmannian coordinates
of the ${\cal N}=2$ superspace is zero and $R=N+\frac12B$. The
supercharge which is manifest in ${\cal N}=2$ superspace formalism
has $R$-charge one, whereas a nonmanifest supercharge has
vanishing $R$-charge.

Let us consider topology-changing operators which belong to ${\cal
N}=4$ (Anti)BPS multiplets.  In the IR limit the theory flows to
the interacting superconformal field theory and (Anti)BPS
representations are labelled by the (anti-)chiral primary
operators.  The conformal dimensions of (anti-)chiral primary
operators are smaller than those of other operators in the same
representation and are determined by their spin and $R$-symmetry
representations \cite{GIOS}-\cite{M}. We define an (Anti)BPS
monopole operator as a topology-changing operator which is an
(anti-)chiral operator with a lowest conformal weight among the
(anti-)chiral topology-changing operators with a given magnetic
charge $H$. It follows that (Anti)BPS monopole operators are
(anti-)chiral primaries. Using arguments similar to those
presented in section~\ref{SYM} we conclude that in the large $N_f$
limit we have matter fields in a background of  (Anti)BPS
monopole.
 Our goal will be to determine the quantum numbers of (Anti)BPS
monopole operators in the limit of large $N_f$.

 Now we will identify (Anti)BPS backgrounds corresponding to
(Anti)BPS GNO monopoles in ${\cal N}=4$ supersymmetric gauge
theory. Background values of $\vec{V}^\alpha$, $\phi^\alpha$,
$\phi^{*\alpha }$, and $\chi^\alpha$ preserve some of the manifest
${\cal N}=2$ supersymmetry parameterized by $\xi,\bar{\xi}$ iff
they satisfy the equations \be\label{susy_1}
\delta\lambda^\alpha=-i\lb\sigma^i\lb\partial_i
\chi^\alpha+f^{\alpha\beta\gamma}\chi^\beta
V^\gamma_i\rb+\frac{1}{2}\epsilon^{ijk}\sigma^k
V^\alpha_{ij}-D^\alpha\rb\xi=0, \ee \be\label{susy_2}
\delta\bar{\lambda}^\alpha=-i\bar{\xi}\lb\sigma^i\lb\partial_i
\chi^\alpha+f^{\alpha\beta\gamma}\chi^\beta
V^\gamma_i\rb-\frac{1}{2}\epsilon^{ijk}\sigma^k
V^\alpha_{ij}+D^\alpha\rb =0, \ee \be\label{susy_3}
\delta\eta^\alpha=\sqrt{2}\lb
f^{\alpha\beta\gamma}\chi^\beta\phi^\gamma+i\sigma^i\lb\partial_i\phi^\alpha+f^{\alpha\beta\gamma}\phi^\beta
V^\gamma_i\rb\rb\bar{\xi}+\sqrt{2}\xi K^\alpha=0, \ee
\be\label{susy_4} \delta\bar{\eta}^\alpha=-\sqrt{2}\xi\lb
f^{\alpha\beta\gamma}\chi^\beta\phi^{*\gamma}+i\sigma^i\lb\partial_i\phi^{*\alpha
}+f^{\alpha\beta\gamma}\phi^{*\beta}V^\gamma_i\rb\rb+\sqrt{2}\bar{\xi}
K^{*\alpha }=0. \ee The other set of supersymmetry transformations
is obtained from (\ref{susy_1})-(\ref{susy_4}) by the replacements
$\lambda\to\eta$, $\eta\to-\lambda$. Consider a background with
$\Phi=0$. Let us set $D^\alpha=0$ and introduce
$E^\alpha_i=-\partial_i
\chi^\alpha-f^{\alpha\beta\gamma}\chi^\beta V^\gamma_i$,
$B^{\alpha i} =\frac12 \epsilon^{ijk}V^\alpha_{jk}$. Equations
(\ref{susy_1})-(\ref{susy_4}) imply
$$
(\vec{E}^\alpha-\vec{B}^\alpha)\vec{\sigma}\xi=0,\quad
\bar{\xi}(\vec{E}^\alpha+\vec{B}^\alpha)\vec{\sigma}=0.
$$
For $\vec{B}=\vec{B}^\alpha T^\alpha=\frac{H}{r^3}\vec{r}$ we have
the following backgrounds, each preserving half of the manifest
${\cal N}=2$
supersymmetry:\\
(i) BPS background
$$
 \vec{E}^\alpha=-\vec{B}^\alpha, \quad \forall\bar{\xi},\quad \xi=0,
$$
(ii) AntiBPS background
$$
\vec{E}^\alpha=\vec{B}^\alpha, \quad \forall\xi,\quad \bar{\xi}=0.
$$
We choose $\chi=\chi^\alpha T^\alpha=\mp H/r$ with
$H=qT^3=\frac12(q,-q)$, where upper (lower)  sign corresponds to
the (Anti)BPS monopole backgrounds\footnote{We will use this
convention throughout the paper.}. These backgrounds are invariant
under $SU(2)_R$ symmetry, break ${\cal N}=4$ to ${\cal N}=2$
supersymmetry, $SU(2)_{gauge}$ group to $U(1)_{gauge}$ subgroup,
and  $SU(2)_N$ to $U(1)_N$. We mention that contrary to monopoles
in $U(1)$ gauge theory, $SU(2)$ monopoles specified by $H$ and
$-H$ are gauge equivalent.

\subsection{\large \bf Dual Theory}

The dual theory is a twisted ${\cal N}=4$,
$\left[U(2)^{N_f-3}\times U(1)^4\right]/U(1)_{diag}$ gauge theory
based on the Dynkin diagram of $SO(2N_f)$ group. The fields
include $N_f-3$ $U(2)$ vector superfields which are made of ${\cal
N}=2$ $U(1)$ vector superfields $U_l$ and neutral chiral
superfields $T_l$, $SU(2)$ vector superfields $T_l$ and adjoint
chiral superfields $S_l$, $l=1,\dots,N_f-3$. Also, there are four
additional $U(1)$ vector superfields which consist of ${\cal N}=2$
vector superfields $U_{N_f-2}$,..., $U_{N_f+1}$ and neutral chiral
superfields $T_{N_f-2}$,..., $T_{N_f+1}$. Factorization of the
diagonal $U(1)$  implies the constraints
$$
\sum_{p=1}^{N_f+1}U_p=0,\quad \sum_{p=0}^{N_f+1} T_p=0.
$$
Matter fields include twisted $N_f-4$ matter hypermultiplets made
of ${\cal N}=2$ chiral multiplets $q_r$ and $\tilde{q}_r$,
transforming as
$$
q_r\to U(2)_{r+1}q_r U(2)^+_r,\quad \tilde{q}_r\to
U(2)_r\tilde{q}_r U(2)^+_{r+1},\qquad r=1,\dots, N_f-4.
$$
  We also have four additional twisted matter hypermultiplets which
decompose with respect to ${\cal N}=2$ as chiral superfields $\lb
X_1,\tilde{X}_1\rb$, ..., $\lb X_4,\tilde{X}_4\rb$. $X_1$
$\lb\tilde{X}_1\rb$ has charge $+1$ ($-1$) under $U(1)_{N_f-2}$
and transforms according to (anti-)fundamental representation of
$U(2)_1$;  $X_2$ $\lb\tilde{X}_2\rb$ has $U(1)_{N_f-1}$ charge
$+1$ ($-1$) and is belongs to (anti-)fundamental representation of
$U(2)_{N_f-3}$; $X_3$ $\lb\tilde{X}_3\rb$ has a charge $+1$ ($-1$)
under $U(1)_{N_f}$ and transforms according to (anti-)fundamental
representation of $U(2)_1$; $X_4$ $\lb\tilde{X}_4\rb$ has a charge
$+1$ ($-1$) under $U(1)_{N_f+1}$ and is transformed according to
(anti-)fundamental representation of $U(2)_{N_f-3}$.
Superpotential is given by
$$
W = i\sqrt{2}\left\{ \tilde{X_1}\lb T_1+S_1-T_{N_f-2}\rb X_1+
\tilde{X}_2\lb T_{N_f-3}+S_{N_f-3}-T_{N_f-1}\rb
X_2+\phantom{\sum_{n}^{N_f}}\right.
$$
$$
+\tilde{X_3}\lb T_1+S_1-T_{N_f}\rb X_3+ \tilde{X_4}\lb
T_{N_f-3}+S_{N_f-3}-T_{N_f+1}\rb X_4+
$$
$$
\left.+\sum_{r=1}^{N_f-4}\tilde{q}_r\lb S_{r+1}+T_{r+1}-S_r-T_r\rb
q_r \right\}.
$$
The two dimensional Higgs branch doesn't receive quantum
corrections and is given by a hyper-Kahler quotient parameterized
by $x$, $y$, and $z$ subject to a constraint
\be\label{hyper-Kahler} x^2+y^2z=z^{N_f-1}. \ee Explicit form of
these coordinates is given in Ref.\cite{LRU}: \be\label{def1}
z=-X^{a_1}_1\tilde{X}_{3|a_1}X^{b_1}_3 \tilde{X}_{1|b_1}, \ee and
(for even $N_f$)
\begin{eqnarray}\label{def2}
x&=&2X^{a_1}_1q^{a_2}_{1|a_1}\dots
q^{a_{N_f-3}}_{N_f-4|a_{N_f-4}}\tilde{X}_{2|a_{N_f-3}}
X^{b_{N_f-3}}_2\tilde{q}^{b_{N_f-4}}_{N_f-4|b_{N_f-3}}\dots
\tilde{q}^{b_1}_{1|b_2}\tilde{X}_{3|b_1}X^{c_1}_3\tilde{X}_{1|c_1},\\
y&=&2X^{a_1}_3q^{a_2}_{1|a_1}\dots
q^{a_{N_f-3}}_{N_f-4|a_{N_f-4}}\tilde{X}_{2|a_{N_f-3}}
X^{b_{N_f-3}}_2\tilde{q}^{b_{N_f-4}}_{N_f-4|b_{N_f-3}}\dots
\tilde{q}^{b_1}_{1|b_2}\tilde{X}_{3|b_1}+(-z)^{N_f/2-1}.\nonumber
\end{eqnarray}
$4N_f-6$ dimensional Coulomb branch is parameterized by $N_f+1$
dual $U(1)$ photons $V_{\pm|r}$ (for a given $r$, $V_{+|r}$ and
$V_{-|r}$ are used as coordinates on two distinct patches) subject
to the constraints
$$
\prod_r V_{+|r}=\prod_r V_{-|r}=1,
$$
 $N_f$ independent chirals $T$, $2N_f-6$ independent coordinates
analogous to the ones given in Eq.(\ref{CB}).

\subsection{\large \bf Mirror Symmetry}

Since mirror symmetry exchanges mass and Fayet-Iliopoulos terms,
we identify $N_f$ complex mass terms $\tilde{Q}^sQ^s$ (no sum over
$s$) with $N_f$ independent chirals $T$. Therefore chirals $T$ and
$S$ have baryon charge $2$ whereas baryon charges of $X$,
$\tilde{X}$, $q$, and $\tilde{q}$ are $-1$. Baryon charges of $x$,
$y$, and $z$ are $2-2N_f$, $4-2N_f$, and $-4$ respectively which
can be deduced both from the defining equations
(\ref{def1})-(\ref{def2}) and from the hyper-Kahler quotient
equation (\ref{hyper-Kahler}). Likewise, $T$ and $S$  have
vanishing $U(1)_N$ charges, whereas $X$, $\tilde{X}$, $q$, and
$\tilde{q}$ have a charge $+1$. Finally, $U(1)_N$ charges of $x$,
$y$, and $z$ are $2N_f-2$, $2N_f-4$, and $4$ respectively. We also
have $R(x)=N_f-1$, $R(y)=N_f-2$, as well as $R(z)=2$.

It follows that charges of $z$ are independent of $N_f$ and
coincide with that of $w=2\textrm{ Tr }\Phi^2$.  Also comparing
Eq.(\ref{classCoulomb}) with Eq.(\ref{hyper-Kahler}) we obtain an
identification
$$
u\sim x,\quad v\sim y,\qquad w\sim z.
$$
Thus, the mirror symmetry predicts the following charges for
operators defined in Eq.(\ref{CB})
\begin{displaymath}
\begin{tabular}{c|c|c|c}
& $N$ & $B$ & $R$\\
\hline
$u$ & $2N_f-2$ & $2-2N_f$ & $N_f-1$\\
\hline
$v$ & $2N_f-4$ & $4-2N_f$ & $N_f-2$\\
\hline
$w$ & $4$ & $-4$ & $2$\\
\end{tabular}
\end{displaymath}
Since $x$, $y$, and $z$ are chiral primary operators which are
polynomials of the electrically charged fields, operators $u$,
$v$, and $w$ are also chiral primaries and describe the sector
with nontrivial magnetic charge. As explained in Ref.\cite{BKW2},
the conformal dimension of ${\cal N}=4$ (anti-)chiral primary
operator equals (minus) the corresponding $U(1)_R$ charge.

\subsection{\large \bf Quantum Numbers}\label{QuantumNumbers}

Quantum numbers of the (Anti)BPS monopole state receive
contributions from both matter hypermultiplet ${\cal Q}$ and
vector multiplet ${\cal V}$. The former is proportional to $N_f$
and is dominant in the large $N_f$ limit, whereas the latter gives
correction of the form $O(1)$. Let us determine the matter
contributions first.

Energy spectra of matter fields in (Anti)BPS backgrounds are given
in the Appendix. Since matter fermionic particles and
antiparticles have different energy spectra we adopt the
``symmetric '' ordering for the bilinear fermionic observables
\be\label{symordering} \bar{\psi}O\psi\to
\frac12{\bar{\psi}}O{\psi}- \frac12{\psi}O^T{\bar{\psi}}, \ee
where $O$ is some operator independent of the fields. To regulate
expression on the RHS of Eq.(\ref{symordering}) we use a
substraction technique. This procedure gives
$$
E^{\textrm{Fermions}}_{\textrm{Casimir}}=\frac12\lb\sum E^--\sum
E^+ \rb-"\textrm{the same}"|_{q=0},
$$
where ${E^+}$, ${E^-}$ are positive and negative energies
respectively. To define the formal sums appearing in this section
we use
$$
\sum E\to\sum Ee^{-\beta|E|}
$$
regularization and take $\beta \to 0$ limit at the end of
calculations.
 Matter bosonic particles and antiparticles have identical energy
spectra and standard prescription can be used. In our model the
matter contribution to the Casimir energy is equal to $h_{{\cal
Q}}=N_f|q|$ for both BPS and AntiBPS monopole backgrounds. For
matter part of the $R$-charge operator we have
$$
R_{{\cal Q}}=\frac14\left[ \sum (a^+_\psi a_\psi-a_\psi a^+_\psi)+
\sum(b_\psi b^+_\psi-b^+_\psi b_\psi)+
 \sum (a^+_{\tilde{\psi}} a_{\tilde{\psi}}-a_{\tilde{\psi}}
a^+_{\tilde{\psi}})+
 \right.
$$
$$
\left.\sum(b_{\tilde{\psi}} b^+_{\tilde{\psi}}-b^+_{\tilde{\psi}}
b_{\tilde{\psi}})-\sum(a^+_A a_A+b_A b^+_A) -\sum(a^+_{\tilde{A}}
a_{\tilde{A}}+b_{\tilde{A}} b^+_{\tilde{A}}) \right]+const,
$$
where $a^+$ ($b^+$) and $a$ ($b$) denote the corresponding
(anti-)particle creation and annihilation operators respectively.
To fix a constant we define a vacuum state with zero magnetic
charge $\left|0\right>$ to have vanishing $R$-charge. It follows
that
$$
\left<R_{{\cal Q}}\right>_{q}=\left( \sum_{E^-_\psi}|E^-_\psi|-
\sum_{E^+_\psi}E^+_\psi+\sum_{E^+_A}E^+_A-
\sum_{E^-_A}|E^-_A|\right)-''\textrm{the same}''|_{q=0}.
$$
As a result we have $\left<R_{{\cal Q}}\right>=\pm h_{{\cal Q}}$.
For $N$ and $B$ charges similar calculations give $\left<N_{{\cal
Q}}\right>_{q}=-\left<B_{{\cal Q}}\right>_{q}=\pm 2N_f|q|$.

Now we will consider the vector multiplet contribution to the
quantum numbers of the vacuum state. Relevant charges are
summarized in the table
\begin{displaymath}
\begin{tabular}{c|c|c|c}
& $N$ & $B$ & $R$\\
\hline
$\lambda$ & $1$ & $0$ & $1$\\
\hline $\phantom{}             $ & $\phantom{}  $ & $\phantom{} $
&
$\phantom{}$\\
$\bar{\lambda}$ & $-1$ & $0$ & $-1$\\
\hline
$\eta$ & $1$ & $-2$ & $0$\\
\hline
$\bar{\eta}$ & $-1$ & $2$ & $0$\\
\hline
$\phi$ & $2$ & $-2$ & $1$\\
\hline
$\phi^*$ & $-2$ & $2$ & $-1$\\
\hline
$\chi$ & $0$ & $0$ & $0$\\
 \end{tabular}
\end{displaymath}
Integration over the hypermultiplet ${\cal Q}$ produces an induced
action for the vector multiplet $S_{Ind}[{\cal V}]$ proportional
to $N_f$. Let us assume that supersymmetric monopole configuration
minimizes vector multiplet effective action in the IR region.
Changing ${\cal V}\to{\cal V}^{mon}+\hat{{\cal V}}/\sqrt{N_f}$
gives
$$
S_{Ind}[{\cal V}]=S^{(2)}_{Ind}[\hat{{\cal
V}}]+O\lb\frac{1}{\sqrt{N_f}}\rb,
$$
where $S^{(2)}_{Ind}[\hat{{\cal V}}]$ is quadratic in $\hat{{\cal
V}}$ and independent from $N_f$. The full effective action for
$\hat{{\cal V}}$ is
$$
S_{Eff}[{\cal V}]=\frac{S_0[{\cal V}]}{e^2}+S_{Ind}[{\cal
V}]=N_f\frac{S_0[{\cal
V}^{mon}]}{\hat{e}^2}+\frac{S^{(2)}_0[\hat{\cal V}]}{\hat{e}^2}+
S^{(2)}_{Ind}[\hat{{\cal V}}]+O\lb\frac{1}{\sqrt{N_f}}\rb,
$$
where $S_0$ is original action for a vector superfield and
$\hat{e}^2=e^2N_f$. Linear term proportional to $\frac{\delta
S_0}{\delta {\cal V}}\left[{\cal V}^{mon}\right]\hat{{\cal V}}$
vanishes because supersymmetric field configuration ${\cal
V}^{mon}$ automatically minimizes the action $S_0$.

The superconformal algebra arising in the IR limit has generators
${\mathbb S}$ and $\bar{\mathbb S}$ which are superpartners of the
special conformal transformations $\mathbb K$:
$$
\left[{\mathbb K}, {\mathbb Q}\right]\sim \bar{\mathbb S},\quad
\left[{\mathbb K}, \bar{\mathbb Q}\right]\sim {\mathbb S}.
$$
In the Eucledean space we have ${\mathbb Q}^+={\mathbb S}$ and
$\bar{\mathbb Q}^+=\bar{\mathbb S}$, hence special conformal
transformations generated by $\bar{\mathbb S}$ $\lb{\mathbb S}\rb$
leave the (Anti)BPS background invariant. The (Anti)BPS background
breaks some of the global symmetries, and the full Hilbert space
of states is given by a tensor product of the physical Fock space
constructed from a vacuum state and a space of superfunctions on
the appropriate supercoset \cite{BKW2}. In Ref.\cite{M} it was
shown that unitarity condition applied to the anticommutator
$\{\bar{\mathbb Q},\bar{\mathbb S}\}$ in the ${\cal N}=2$
supersymmetric theory implies that the  conformal weight $h$ and
infrared $R$-charge $R_{IR}$ of any state satisfy $h\ge R_{IR}$.
Also, it follows from the anticommutator $\{{\mathbb Q},{\mathbb
S}\}$ that $h\ge-R_{IR}$ in a unitary theory. As explained in
Ref.\cite{BKW2}, the infrared $R$-charge $R_{IR}$ coincides with
$U(1)_R$ charge $R$. Thus for the physical Fock space constructed
from the (Anti)BPS vacuum, the unbroken subalgebra of the
three-dimensional superconformal algebra implies that (minus)
$R$-charge of a state can not exceed its conformal dimension:
\be\label{UnitarityBound} h\ge\pm R, \ee with the lower bound
saturated by the (anti-)chiral primary operators with vanishing
spin.

 Let us focus on the gluino contribution to the $R$-charge.
We have two sets of gluinos $\hat{\lambda}$ and  $\hat{\eta}$.
Since $U(1)_R$ symmetry acts trivially on $\hat{\eta}$, the only
contribution comes from $\hat{\lambda}$. Relevant quadratic terms
in the effective action have the form (in $R^3$):
$$
S^{(2)}_0[\hat{\lambda},\hat{\bar{\lambda}}]=\int\phantom{} dx
\lb{i\hat{\bar{\lambda}}_+\left[\vec{\sigma}\lb\vec{\nabla}
-i\vec{V}^{mon}\rb\pm\frac{q}{r}\right]\hat{\lambda}_+}+
{i\hat{\bar{\lambda}}_-\left[\vec{\sigma}\lb\vec{\nabla}
+i\vec{V}\rb\mp\frac{q}{r}\right]\hat{\lambda}_-}+
{i\hat{\bar{\lambda}}^3\vec{\sigma}\vec{\nabla} \hat{\lambda}^3}
\rb,
$$
$$
S^{(2)}_{Ind}[\hat{\lambda},\hat{\bar{\lambda}}]=\int\phantom{}dxdy\lb\hat{\bar{\lambda}}_+(x)O^{(+)}(x,y)\hat{\lambda}_+(y)+
\hat{\bar{\lambda}}_-(x)O^{(-)}(x,y)\hat{\lambda}_-(y) \right.
$$
$$
+\left.\hat{\bar{\lambda}}^3(x)O^{(3)}(x,y)\hat{\lambda}^3(y)\rb,
$$
with
$\hat{\lambda}_+=(\hat{\lambda}^1+i\hat{\lambda}^2)/\sqrt{2}$,
$\hat{\lambda}_-=(\hat{\lambda}^1-i\hat{\lambda}^2)/\sqrt{2}$, and
$$
O^{(+)}(x,y)\sim\left<\bar{\psi}_1(x)\psi_1(y)\right>\left<A_2(x)A_2^+(y)\right>,
\quad
O^{(-)}(x,y)\sim\left<\bar{\psi}_2(x)\psi_2(y)\right>\left<A_1(x)
A_1^+(y)\right>,
$$
$$
O^{(3)}(x,y)\sim\left<\bar{\psi}_1(x)\psi_1(y)\right>\left<A_1(x)A_1^+(y)\right>+\left<\bar{\psi}_2(x)\psi_2(y)\right>\left<A_2(x)A_2^+(y)\right>,
$$
where we used the identities
$$
\left<\bar{\tilde{\psi}}_1(x)\tilde{\psi}_1(y)\right>=
\left<\bar{\psi}_2(x)\psi_2(y)\right>,\quad
\left<\bar{\tilde{\psi}}_2(x)\tilde{\psi}_2(y)\right>=
\left<\bar{\psi}_1(x)\psi_1(y)\right>,
$$
$$
\left<\tilde{A}_1(x)\tilde{A}^+_1(y)\right>=\left<A_2(x)A^+_2(y)
\right>,\quad
\left<\tilde{A}_2(x)\tilde{A}^+_2(y)\right>=\left<A_1(x)A^+_1(y)
\right>.
$$
$R$-charge contribution of $\hat{\lambda}_+$ and
$\hat{\bar{\lambda}}_+$ can be expressed in terms of
$\mathbb{\eta}$-invariant of the Hamiltonian associated with
$O^{(+)}$. If $\hat{\lambda}_+$ has zero-energy modes in the Fock
space, it may lead to ambiguities in the $R$-charge computation.
Let us show that such modes are not present. Induced action
equation of motion $\delta S^{(2)}_{Ind}/\delta
\hat{\bar{\lambda}}_+=0$ has the form
 \be\label{gluinoEOM}
\int\phantom{}dy O^{(+)}(x,y)\hat{\lambda}_+(y)=0. \ee
Transforming to $S^2\times R$ and assuming $\hat{\lambda}_+$
independent of $\tau$, we obtain
\be\label{gluinoEOMinfinitecoupling} \int\phantom{}d\tau_y
d\varphi_y d\theta_y
O^{(+)}(\varphi_x,\theta_x,\tau_x;\varphi_y,\theta_y,\tau_y)\hat{\lambda}_+(\varphi_y,\theta_y)=0.
\ee If Eq.(\ref{gluinoEOMinfinitecoupling}) has a non-trivial
solution corresponding to an operator acting in the Fock space,
$SU(2)_R$ symmetry implies that
$\hat{\eta}_+=\lb\hat{\eta}_1+i\hat{\eta}_2 \rb/\sqrt{2}$ also has
a zero-energy mode. Then it follows from the supersymmetry
transformation
$$
\delta
\hat{\phi}_+^{*}=\sqrt{2}\bar{\xi}\hat{\bar{\eta}}_+e^{-\tau/2},
$$
 that $\hat{\phi}_+$ has a mode with energy $-1/2$ in the Fock
space associated with BPS monopole background. Let us denote the
corresponding creation operator as
$b_{\hat{\phi}_+}^{+\{|E^-|=1/2\}}$.
 Using the explicit form of the matter field
energy modes it is straightforward to check that
$$
O^{(-)}= O^{(+)}|_{\varphi_x\to-\varphi_x,\varphi_y\to-\varphi_y},
$$
which implies that there is a zero-energy solution for
$\hat{\lambda}_-$ as well. Hence, $\hat{\eta}_-$ has zero-energy
mode and $\hat{\phi}_-$ has a mode with energy $-1/2$ which we
denote as $b_{\hat{\phi}_-}^{+\{|E^-|=1/2\}}$. The product
$b_{\hat{\phi}_+}^{+\{|E^-|=1/2\}}b_{\hat{\phi}_-}^{+\{|E^-|=1/2\}}$
is $U(1)_{gauge}$ invariant operator which  has $R$-charge $2$ and
energy (conformal dimension) $1$. Repeated action of this operator
on any physical state with definite $R$-charge and conformal
dimension will finally give a state with $R$-charge greater than
the conformal dimension which violates the unitarity bound
(\ref{UnitarityBound}). Thus we conclude that $\hat{\lambda}_+$
does not have zero-energy modes in the Fock space constructed from
the BPS vacuum. For AntiBPS monopole background similar analysis
gives analogous conclusion.

Action $S^{(2)}_{Eff}$ acquires explicit $\tau$-dependence on
$S^2\times R$ as a reminiscence of the fact that theory is not
conformal invariant for $0<\hat{e}^2<\infty$. Let us define
$g^2=e^\tau\hat{e}^2$ and consider the resulting theory
$S^{(2)}_{g}$, which can be viewed as conformal invariant
deformation of $S^{(2)}_{Ind}$ with constant $g$ being a
deformation parameter.
 Let $\{E_k(g)\}$ be the energy spectrum of
$\hat{\lambda}_+$, then the $R$-charge contribution is
\be\label{R_gluinos}
<R_{\hat{\lambda}_+,\hat{\bar{\lambda}}_+}>_{q}=\lim_{\beta\to0}
\lb Z(q,\beta)-Z(0,\beta) \rb,\quad Z(q,\beta)=\frac12\sum_k
\textrm{sign}\left[ E_k\lb g\rb\right] e^{-\beta \left| E_k\lb
g\rb\right|} . \ee
Since $Z(q,\beta)$ is proportional to $\mathbb{\eta}$-invariant,
$<R_{\hat{\lambda}_+,\hat{\bar{\lambda}}_+}>_q$ is expected to be
independent from $g$ and, hence, can be computed in the region of
small $g$. To make this argument rigorous it is necessary to show
that $\hat{\lambda}_+$ does not have zero-energy modes for all
values of the constant $g$. We hope to return to this problem in
the future. If $g$ is small the induced action terms can be
ignored and we have gluinos moving in a monopole background ${\cal
V}^{mon}$. The Hamiltonian eigen-value equation has a form
$$
\lb \left. {\cal
H}_{\psi_2}\right|_{q\to2q}-\frac12\rb\hat{\lambda}_+=E\hat{\lambda}_+,
$$
 where ${\cal H}_{\psi_2}$ is Hamiltonian for the matter field
$\psi_2$. Using energy spectrum of $\psi_2$ given in the Appendix,
we find that the spectrum of $\hat{\lambda}_+$ in the limit of
small $g$ is given by $(n=1,2,\dots)$
$$
E = -|q|-n-\frac12, \quad \mp|q|-\frac12,\qquad |q|+n-\frac12,
$$
where each energy level has degeneracy $|2E+1|$. We mention that
energy level $E=\mp|q|-\frac12$ has degeneracy $2|q|$ and is not
present if $q=0$.
  Using Eq.(\ref{R_gluinos}) in the small $g$ region we obtain
$<R_{\hat{\lambda}_+,\hat{\bar{\lambda}}_+}>_q=\mp|q|$.

Similar analysis can be implemented for $\hat{\lambda}_-$ and
$\hat{\lambda}^3$. $R$-charge contribution of $\hat{\lambda}_-$ is
identical to that of $\hat{\lambda}_+$, whereas in the small $g$
limit $\hat{\lambda}^3$ is moving in the trivial $({\cal V}=0)$
background and does not contribute to the $R$-charge. Besides
gluinos $\hat{\lambda}$, the only vector multiplet fields charged
under the $U(1)_R$ symmetry are scalars $\phi$ and $\phi^*$.
Analogous calculations show that they do not contribute to the
$O(1)$ terms of the $R$-charge. Therefore, in the large $N_f$
limit, we have
$$
\left<R\right>_{q}=\pm\lb N_f-2\rb|q|.
$$
For the $N$-charge we have $\left<N_{{\cal V}}\right>_q=
<N_{\hat{\lambda},\hat{\bar{\lambda}}}>_q+<N_{\hat{\eta},\hat{\bar{\eta}}}>_q
=2<N_{\hat{\lambda},\hat{\bar{\lambda}}}>_q, $ where we used
invariance of the (Anti)BPS background under $SU(2)_R$:
$$
S^{(2)}_{Ind}[\hat{\eta},\hat{\bar{\eta}}]=\left.
S^{(2)}_{Ind}[\hat{\lambda},\hat{\bar{\lambda}}]\right|_{\hat{\lambda}\to\hat{\eta},\hat{\bar{\lambda}}\to\hat{\bar{\eta}}}.
$$
Calculations similar to those for the $R$-charge give
$\left<N_{\cal V}\right>_q=\mp4|q|$, which implies
$$
\left<N\right>_q=\pm\lb 2N_f-4 \rb|q|,\quad
\left<B\right>_q=\pm(4-2N_f)|q|.
$$

\subsection{\large\bf Comparison with the Mirror Symmetry Predictions}

Mirror symmetry implications for the quantum numbers of $w$ (see
Eq.(\ref{CB})) are trivially satisfied. Thus we conclude that
$w\sim z$. Let us consider a physical state $\left|vac\right>_q$.
It is the lowest energy state and, therefore, a superconformal
primary. Since (Anti)BPS background is annihilated by a
supercharge $\bar{\mathbb{Q}}$ $\lb\mathbb{Q}\rb$, a state
$\bar{\mathbb{Q}}\left|vac\right>_q$
$\left(\mathbb{Q}\left|vac\right>_q\right)$ belongs to the Fock
space associated with $\left|vac\right>_q$. Since supercharge
$\bar{\mathbb{Q}}$ $\lb\mathbb{Q}\rb$ raises energy by $1/2$ and
has $U(1)_R$ charge (minus) one, we find that there is no such a
physical state in the Fock space. Therefore, a state
$\left|vac\right>_q$ is annihilated by $\bar{\mathbb{Q}}$
$\lb\mathbb{Q}\rb$ and corresponds to the insertion of the
(anti-)chiral primary operator at the origin of $R^3$. Thus,
conformal weight of $\left|vac\right>_q$ equals its $R$-charge,
i.e., $\pm\lb N_f-2\rb|q|$. Background $\Phi=0$ also corresponds
to the (Anti)BPS monopoles in ${\cal N}=2$ $SU(2)$ gauge theory
which can be obtained by giving mass to the adjoint chiral field
$\Phi$. Therefore, matter contribution to the quantum numbers of
the ${\cal N}=2$ (Anti)BPS monopoles is the same as in the ${\cal
N}=4$ theory. We also note that chiral primaries $u$ and $w$ are
present for ${\cal N}=4$ only and absent in ${\cal N}=2$ theory.
This observation implies
$$
\left|vac\right>^{\textrm{BPS}}_{|q|=1}\propto
v(0)\left|0\right>,\quad
\left|vac\right>^{\textrm{AntiBPS}}_{|q|=1}\propto
v^+(0)\left|0\right>.
$$
Identity of $\left|vac\right>^{\textrm{BPS}}_{|q|=1}$ and $y$
quantum numbers gives $v\sim y$.

To obtain another (anti-)chiral primary state in the physical Fock
space, we must act on a state
$\left|vac\right>^{\textrm{(Anti)BPS}}_{|q|=1}$ with an
$U(1)_{gauge}$ invariant operator $f$ such that it raises energy
by $R(f)$ $\lb-R(f)\rb$. It is easy to see that $f$ can not be
made of matter fields only. Indeed, the most general expression
for (anti-)chiral primary $f\lb{\cal Q}\rb\left|vac\right>_q$
would be a superposition of gauge invariant states of the form
$\lb a^+_{\cal Q}\rb^m\lb b^+_{\cal Q}\rb^p\left|vac\right>_q$
with some non-negative integers $m$ and $p$. However,
$$
E\lb a^+_{\cal Q}\rb>\pm R\lb a^+_{\cal Q}\rb,\quad E\lb b^+_{\cal
Q}\rb>\pm R\lb b^+_{\cal Q}\rb,
$$
and the state $\lb a^+_{\cal Q}\rb^m\lb b^+_{\cal
Q}\rb^p\left|vac\right>_q$ is not an (anti-)chiral primary, unless
$m=p=0$.

Now we consider energy spectra of fields which belong to the
vector multiplet. In the IR limit the only terms in the vector
multiplet effective action are those induced by integration over
the matter hypermultiplets.
Let us show that gluinos $\hat{\eta}$, $\hat{\bar{\eta}}$,
$\hat{\lambda}$, $\hat{\bar{\lambda}}$ do not have (anti-)chiral
primary creation operators in the Fock space associated with the
(Anti)BPS background. It follows from Eq.(\ref{UnitarityBound})
that such modes can not be present in $\hat{\bar{\eta}}$ and
$\hat{\bar{\lambda}}$, ($\hat{\eta}$ and $\hat{\lambda}$). Since
$R$-charge of $\hat{\eta}$  vanishes, the (anti-)chiral primary
creation operator corresponds to a mode with zero energy. It was
shown in section~\ref{QuantumNumbers} that $\hat{\eta}_+$ and
$\hat{\eta}_-$, ($\hat{\bar{\eta}}_+$ and $\hat{\bar{\eta}}_-$),
do not have zero-energy modes. If a gauge invariant field
$\hat{\eta}^{(3)}$ $(\hat{\bar{\eta}}^{(3)})$ has a creation
operator with zero energy, then $SU(2)_R$ symmetry implies
existence of $b^{+\{E=0\}}_{\hat{\lambda}^{(3)}}$,
($a^{+\{E=0\}}_{\hat{\lambda}^{(3)}}$), which is incompatible with
Eq.({\ref{UnitarityBound}}).
 If present, an (anti-)chiral primary creation operator of
$\hat{\lambda}^\alpha$ ($\hat{\bar{\lambda}}^\alpha$)  has the
form $b^{+\{|E^-|=1\}}_{\hat{\lambda}^\alpha}$,
($a^{+\{|E^-|=1\}}_{\hat{\lambda}^\alpha}$). Then $SU(2)_R$
symmetry ensures existence of
$b^{+\{|E^-|=1\}}_{\hat{\eta}^\alpha}$,
($a^{+\{|E^-|=1\}}_{\hat{\eta}^\alpha}$). The supersymmetry
transformation
$\delta\hat{\phi}^{*\alpha}=\sqrt{2}\bar{\xi}\hat{\bar{\eta}}^\alpha
e^{-\tau/2}$,
($\delta\hat{\phi}^{\alpha}=\sqrt{2}{\xi}\hat{{\eta}}^\alpha
e^{\tau/2}$), implies a presence of $\hat{\phi}^\alpha$ mode with
energy $|E^-|=3/2$: $\hat{\phi}^\alpha \sim e^{3\tau/2}$,
$\hat{\phi}^{*\alpha} \sim e^{-3\tau/2}$.
Such modes should annihilate the right-hand-side of $S^2\times R$
counterpart
 of Eq.(\ref{susy_3}), (Eq.(\ref{susy_4})), for all $\bar{\xi}$
$(\xi)$ in the (Anti)BPS monopole background
 to ensure that an operator $b^{+\{|E^-|=1\}}_{\hat{\eta}^\alpha}$,
($a^{+\{|E^-|=1\}}_{\hat{\eta}^\alpha}$) is annihilated by
$\bar{\mathbb{Q}}$ ($\mathbb{Q}$).
 It is
easy to see that it can not be the case. Thus we conclude that
gluinos do not have (anti-)chiral primary creation operators in
the Fock space. Similar arguments reveal that it is true for
$\hat{\chi}^\alpha$ and $\hat{V}^\alpha_i$ as well. Thus
$\hat{\phi}$ and $\hat{\phi}^*$ are the only fields which could
have such modes.

It follows from Eq.(\ref{UnitarityBound}) that energy spectrum of
$\hat{\phi}^\alpha$ satisfies $|E^-|\ge
R\lb\hat{\phi}^\alpha\rb=1$, $\lb E^+\ge
-R\lb\hat{\phi}^{*\alpha}\rb=1\rb$. The (Anti)BPS background under
consideration has vanishing expectation values of $U(1)_{gauge}$
invariant fields $\phi^{(3)}$ and $\phi^{*(3)}$. However, as it
follows from Eqs.(\ref{susy_1})-(\ref{susy_4}), setting
$\phi^{(3)}=c$, $\lb\phi^{*(3)}=c\rb$\footnote{It also implies
setting $\phi^{*(3)}$ $\lb\phi^{(3)}\rb$ to $c^*/r^2$ in $R^3$.},
with constant $c$ in $R^3$ leaves the (Anti)BPS background
invariant under $\mathbb{\bar{Q}}$ $\lb\mathbb{Q}\rb$. Therefore,
the action $S_{Eff}[{\cal V}]$ is stationary on these field
configurations. Since the constant $c$ is arbitrary, quadratic
part of $S_{Eff}[\hat{\cal V}]$ is stationary as well.
 In the IR limit it implies existence of the creation operator
$b^{+\{|E^-|=1\}}_{\hat{\phi}^{(3)}}$ $\lb
a^{+\{E^+=1\}}_{\hat{\phi}^{(3)}}\rb$, corresponding to the
spinless mode of $\hat{\phi}^{(3)}$ $\lb\hat{\phi}^{*(3)}\rb$ on
$S^2\times R$. In the (Anti)BPS background any creation operator
of $\hat{\phi}^{(3)}$ $\lb\hat{\phi}^{*(3)}\rb$ corresponding to a
mode with energy $|E^-|=1$ $\lb E^+=1\rb$ saturates the unitarity
bound given by Eq.(\ref{UnitarityBound}). Hence, this mode has
vanishing spin and is given by $const\times e^\tau$ on $S^2\times
R$. Thus the (anti-)chiral primary mode of $\hat{\phi}^{(3)}$
$\lb\hat{\phi}^{*(3)}\rb$ in the (Anti)BPS background is unique.
Acting with the corresponding creation operators on the state
$\left|vac\right>^{\textrm{(Anti)BPS}}_{|q|=1}$ we obtain chiral
primaries with the quantum numbers identical to those predicted
for $u$ $\lb u^+\rb$. We have
$$
b^{+\{|E^-|=1\}}_{\hat{\phi}^{(3)}}\left|vac\right>^{\textrm{BPS}}_{|q|=1}\propto
u(0)\left|0\right>,\quad
a^{+\{E^+=1\}}_{\hat{\phi}^{(3)}}\left|vac\right>^{\textrm{AntiBPS}}_{|q|=1}\propto
u^+(0)\left|0\right>.
$$
The BPS background breaks the Weyl subgroup $Z_2$ spontaneously
and $Z_2$ invariance of the physical states is not required.
However, it might be instructive to construct $Z_2$ invariant
(anti-)chiral primary states by "integrating" the physical states
over $Z_2$. Let us introduce a pair of gauge equivalent states
$$
\left|vac\right>^{\textrm{BPS}}_{q=1}\propto
Y_+\left|0\right>,\quad
\left|vac\right>^{\textrm{BPS}}_{q=-1}\propto Y_-\left|0\right>.
$$
Then,
$$
v(0)\left|0\right>\propto
\left|vac\right>^{\textrm{BPS}}_{q=1}+\left|vac\right>^{\textrm{BPS}}_{q=-1},
\quad u(0)\left|0\right>\propto
\phi^{(3)}\lb\left|vac\right>^{\textrm{BPS}}_{q=1}-\left|vac\right>^{\textrm{BPS}}_{q=-1}\rb.
$$
Similar construction can be made for the AntiBPS monopole
operators as well.

\bigskip

\section{\Large \bf Discussion}\label{Discussion}

We have studied monopole operators in non-supersymmetric $SU(N_c)$
gauge theories as well as (Anti)BPS monopole operators in ${\cal
N}=4$ $SU(2)$ gauge theories in the limit of large $N_f$. In the
case of $SU(N_c)$ non-supersymmetric gauge theories we found that
monopole operators with minimum magnetic charge have zero spin and
transform nontrivially under the flavor symmetry group. Conformal
dimensions of these operators have leading terms of the order
$N_f$ and further sub-leading corrections are expected.

In the case of ${\cal N}=4$ $SU(2)$ gauge theory, the mirror
symmetry predicts existence of two (anti-)chiral primary monopole
operators corresponding to the (anti-)chiral primary operators $x$
$\lb x^+\rb$  and $y$ $\lb y^+\rb$ in the dual theory. The
(anti-)chiral primary operator dual to $y$ $\lb y^+\rb$ exists in
${\cal N}=2$ theory as well, whereas existence of the
(anti-)chiral primary dual to $x$ $\lb x^+\rb$ is a special
feature of ${\cal N}=4$ theory.  Using the radial quantization we
have shown that a state
$\left|vac\right>^{\textrm{(Anti)BPS}}_{|q|=1}$ corresponds to the
insertion of the (anti-)chiral primary monopole operator which is
dual to the operator $y$ $\lb y^+\rb$ in the large $N_f$ limit. We
demonstrated that there is unique (anti-)chiral primary monopole
operator with quantum numbers matching those of $x$ $\lb x^+\rb$.
However we note that the relation in the chiral ring implied by
Eq.(\ref{hyper-Kahler}) remains obscure.

We have shown that (Anti)BPS monopole operators of ${\cal N}=4$
supersymmetric theory are scalars under the $SU(2)_{rot}$ and
transform trivially under the flavor symmetry group.
Transformation properties under the global symmetries have been
computed in the large $N_f$ limit providing a new nontrivial
verification of three-dimensional mirror symmetry. Although we
perform calculations using $1/N_f$ expansion, our result for
quantum numbers of the (Anti)BPS monopole operators are exact and
do not receive $O(1/N_f)$ corrections. The reason is that the
charges which correspond to $U(1)$ subgroups of the compact
$R$-symmetry group must be integral.

It might be interesting to generalize the analysis of the present
paper to the monopole operators of ${\cal N}=4$ $SU(N_c)$ gauge
theories with $N_c>2$.

\bigskip

\section{\Large \bf Appendix:\\ Radial Quantization of
Three-Dimensional
 $\bf {\cal N}=4$ $\bf SU(2)$ Gauge Theory}

We start with ${\cal N}=2$  lagrangian density in four dimensional
Minkowski space\footnote{We adopt the notations of Wess and
Bagger, Ref.\cite{WB}.} for the vector multiplet ${\cal V}$ in the
adjoint representation of $SU(2)$ and hypermultiplets ${\cal Q}^s$
in the fundamental representation of the gauge group
$$
{\cal L}^{R^{3,1}}_{\cal V}=\frac{1}{8e^2}\lb\int\phantom{}
d^2\theta \left.\textrm{Tr}\lb W^\alpha
W_{\alpha}\rb\right|_{\bar{\theta}=0}+h.c.\rb+\frac{1}{e^2}\int\phantom{}
d^2\theta d^2\bar{\theta} \textrm{Tr}(\Phi^+e^{2V}\Phi),
$$
$$
{\cal L}^{R^{3,1}}_{\cal Q}= \int\phantom{} d^2\theta
d^2\bar{\theta} \sum_{s=1}^{N_f}\lb Q^{s+}e^{2V}
Q^s+\tilde{Q}^se^{-2V}\tilde{Q}^{s+}\rb+\lb \int\phantom{}
d^2\theta\left. \phantom{}W\right|_{\bar{\theta}=0}+h.c.\rb,
$$
where a superpotential $W=i\sqrt{2}\sum_{s=1}^{N_f}\tilde{Q}^s\Phi
Q^s$. Let us perform the Wick rotation to $R^4$
$$
{\cal L}_{R^4}=-{\cal L}_{R^{3,1}}|_{x^0=-it}, \quad
V^\alpha_0|_{R^{3,1}}= i\chi^\alpha|_{R^4},
$$
and assume that all fields are independent of the Eucledean time
$t$. This procedure gives ${\cal N}=4$ supersymmetric lagrangian
density in three-dimensional Eucledean space:
$$
{\cal L}_{\cal
Q}^{R^3}=i\bar{\psi}\vec{\sigma}\lb\vec{\nabla}+i\vec{V}\rb\psi+i\bar{\psi}\chi\psi+\lb\left[\vec{\nabla}+i\vec{V}\right]
A\rb^+\lb\left[\vec{\nabla}+i\vec{V}\right] A\rb+A^+\chi^2 A+
$$
$$
i\sqrt{2}\lb \bar{\psi} \bar{\lambda} A- A^+ \lambda\psi\rb
-F^+F-A^+DA+
i\bar{\tilde{\psi}}\vec{\sigma}(\vec{\nabla}-i\vec{V}^T)\tilde{\psi}-i\bar{
\tilde{\psi}}\chi^T\tilde{\psi}+
$$
$$
\lb\left[\vec{\nabla}+i\vec{V}\right]\tilde{A}^+\rb^+\lb\left[\vec{\nabla}+i\vec{V}\right]\tilde{A}^+\rb+\tilde{A}\chi^2\tilde{A}^+-\tilde{F}\tilde{F}^++\tilde{A}D\tilde{A}^+-
i\sqrt{2}\lb\tilde{A}\bar{\lambda}\bar{\tilde{\psi}}-\tilde{\psi}\lambda\tilde{A}^+\rb+\dots,
$$
where the dots denote terms originated from the superpotential and
summations over color and flavor indices are implied. To obtain a
theory on $S^2\times R$ we perform the Weyl rescaling $g_{ij}\to
r^2 g_{ij}$ and introduce $\tau=\ln{r}$. The matter fields
transform as
$$
\lb\psi,\bar{\psi},\tilde{\psi},\bar{\tilde{\psi}}\rb\to
e^{-\tau}\lb\psi,\bar{\psi},\tilde{\psi},\bar{\tilde{\psi}}\rb,
\quad \lb A,A^+,\tilde{A},\tilde{A}^+\rb\to e^{-\frac{\tau}{2}}\lb
A, A^+,\tilde{A},\tilde{A}^+\rb.
$$
For fields in the vector multiplet we have
$$
\lb \chi,\phi,\phi^+ \rb  \to
e^{-\tau}\lb\chi,\phi,\phi^+\rb,\quad \vec{V}\to\vec{V}, \qquad
\lb\lambda, \bar{\lambda},\eta,\bar{\eta}\rb\to
e^{-\frac32\tau}\lb\lambda, \bar{\lambda},\eta,\bar{\eta}\rb.
$$
The (Anti)BPS background is diagonal in color indices and,
therefore, we may use results of Ref.\cite{BKW2} for matter energy
spectra in a background of $U(1)$ monopole with a substitution
$q\to q/2$. Solutions with energy $E$ have the form
$Q,\tilde{Q}\sim e^{-E\tau}$, whereas $Q^+,\tilde{Q}^+\sim
e^{E\tau}$.  To summarize we have, $(n=1,2,\dots)$:
\be\label{SpectrumPsi} E=-\frac{|q|}{2}-n,\quad
\mp\frac{|q|}{2},\qquad \frac{|q|}{2}+n, \ee for $\psi^s_a$,
$\tilde{\psi}^s_a$ and
$$
E=-\frac{|q|}{2}-n,\quad \pm\frac{|q|}{2},\qquad \frac{|q|}{2}+n,
$$
for $\bar{\psi}^s_a$ and $\bar{\tilde{\psi}}^s_a$. Scalar fields
$A^s_a$, $\tilde{A}^s_a$, $A^{+s}_a$, and $\tilde{A}^{+s}_a$ have
$$
E=-\frac{|q|-1}{2}-n,\quad \frac{|q|-1}{2}+n.
$$
Each energy level with energy $E$ has a spin $j=|E|-1/2$ and a
degeneracy $2|E|$. We notice that fermionic spectrum is not
invariant under $E\to-E$. The fact that $A$ and $\tilde{A}^+$ have
identical energy spectra is consistent with the action of
$SU(2)_R$ symmetry. On the other hand fields $\psi$ and
$\bar{\tilde{\psi}}$ have different energy spectra which conforms
with the breaking of $SU(2)_N$ symmetry to a $U(1)_N$ subgroup
which doesn't mix these fermionic fields.

\bigskip

 \centerline{\bf Acknowledgments}

I am grateful to A.~Kapustin for suggesting this project and for
the numerous discussions. I also benefited from conversations with
Xinkai Wu, Yi Li, and Takuya Okuda.

\end{document}